\documentclass[reqno]{amsart} 
\usepackage{enumerate, stmaryrd, amssymb}
\newtheorem{Theorem}{Theorem}
\newtheorem{Proposition}{Proposition}
\newtheorem{Lemma}{Lemma}

\numberwithin{equation}{section}
\newcommand{\yxtt}{|y-x|<t-\tau}
\newcommand{\yxtte}{|y-x|<(1-\epsilon)(t-\tau)}
\newcommand{\yxt}{|y-x|<t}
\newcommand{\pC}{|p|<C(t)}
\newcommand{\sqp}{\sqrt{1+|p|^2}}
\newcommand{\sqttyx}{\sqrt{(t-\tau)^2-|y-x|^2}}
\newcommand{\sqtyx}{\sqrt{t^2-|y-x|^2}}
\newcommand{\sqxi}{\sqrt{1-|\xi|^2}}
\newcommand{\xiDp}{1+\xi \cdot \hat{p}}

\title{Global existence of solutions of the Nordstr\"om-Vlasov system in two space dimensions}
\author{Hayoung Lee}
\address{Max-Planck-institut f\"ur Gravitationsphysik, Am M\"uhlenberg 1, Golm, 14476, Germany}
\email{hayoung@aei.mpg.de}
\begin{document}
\begin{abstract}
The dynamics of a self-gravitating ensemble of collisionless particles is modeled by
the Nordstr\"om-Vlasov system in the framework of the Nordstr\"om scalar theory of gravitation.
For this system in two space dimensions, integral representations of the first order derivatives
of the field are derived. Using these representations we show global existence of smooth solutions
for large data.
\end{abstract}
\maketitle
\markboth{\sc Hayoung Lee}{\sc The Nordstr\"om-Vlasov system in two space dimensions}

\section{Introduction}
The Vlasov equation in general describes a collection of collisionless particles.
Each particle is driven by self-induced fields which are generated by all particles
together. When the relativistic effects are negligible, the dynamics is described by
the Vlasov-Poisson system. Otherwise the relativistic Vlasov-Maxwell system in plasma physics
and the Einstein-Vlasov system in stellar dynamics are considered.
The Vlasov-Poisson models are well understood by now in the question of global existence 
of classical solutions \cite{LP,Pf,R,Sch2}. 
The relativistic models have very different structure and so far they have been considered
separately. In the gravitational case, global existence of (asymptotically flat) 
solutions for the Einstein-Vlasov system is known only for small data with spherical symmetry \cite{RR1}. 
For the relativistic Vlasov-Maxwell system the theory is more developed,
cf. \cite{C2,DL}, \cite{GSh1}--\cite{GS2}, \cite{R2}. 
However global existence and uniqueness of classical solutions 
for large data in three dimensions is still open. 

A different relativistic generalization to the Vlasov-Poisson system 
in the stellar dynamics case has been considered in \cite{C},
where the Vlasov dynamics is coupled to a relativistic scalar theory of gravity 
which goes back, essentially, to Nordstr\"om \cite{No}. 
More precisely, the gravitational theory considered in \cite{C} 
corresponds to a reformulation of Nordstr\"om's theory due to Einstein and Fokker \cite{EF}. 
Therefore the resulting system has been called Nordstr\"om-Vlasov system.

Let $f(t, x, p) \geq 0$ denote the density of the particles in phase space,
where $t \in \mathbb{R}$ denotes time, $x \in \mathbb{R}^2$ position and $p \in \mathbb{R}^2$ momentum. 
The gravitational effects are mediated by a scalar field $\phi(t, x)$.
The Nordstr\"om-Vlasov system in two dimensions is given by
\begin{gather}
\partial^2_t \phi - \Delta_x \phi = - 4 \pi \int f \frac{dp}{\sqrt{1+|p|^2}}\label{NV1}\\
\partial_t f + \hat{p} \cdot \nabla_x f -
\Big[ S(\phi) p + \frac{\nabla_x \phi}{\sqrt{1+|p|^2}}\Big] \cdot \nabla_p f = 3S(\phi)f\label{NV2}
\end{gather}
where $\hat{p} = p(1+|p|^2)^{-1/2}$ and $S= \partial_t + \hat{p}\cdot \nabla_x$.
Initial data are given by 
\begin{align*}
f(0,x,p)&=f^{\rm in}(x,p),\\
\phi(0,x)&=\phi_{0}^{\rm in}(x),\\ 
\partial_t\phi(0,x)&=\phi_1^{\rm in}(x).
\end{align*}

The spacetime is a Lorentzian manifold with a conformally flat metric which,
in the coordinates $(t, x)$, takes the form
$$g_{\mu\nu}= e^{2\phi} {\rm diag}(-1, 1, 1)$$
where the Greek indices run from 0 to 2. The particle distribution $f_{\rm physical}$ defined
on the mass shell in this metric is given by
$$f_{\rm physical}(t, x, p) = e^{-3 \phi} f(t, x, e^{\phi}p).$$ 
Details on the derivation of this system in three dimensions can be founded in \cite{C, CaLe}
and also in general $N$ dimensions in \cite{CaRe}.

In \cite{CaRe} a condition is established such that a global classical solution is achieved in three dimensions
and existence of global weak solutions of the Nordstr\"om-Vlasov system has been shown in \cite{CaRe1}.
Also the Nordstr\"om-Vlasov system has been justified as a genuine relativistic generalization 
of the (gravitational) Vlasov-Poisson system, by indicating the relation between the solutions of the two systems. 
Precisely it has been proved in \cite{CaLe} that in the non-relativistic limit $c\to\infty$
the solutions of the Nordstr\"om-Vlasov system in three dimensional space converge
to solutions of Vlosov-Poisson system in a pointwise sense.
One can prove a similar result in the case of two space dimensions, using the analogous argument
in \cite{LE}.

This paper proceeds as follows. 
In Section \ref{sec:pre_main} we provide representations of the derivatives of the scalar field and state
our main results in detail.
The first of such results is a global existence theorem of solutions of the Nordstr\"om-Vlasov system
under the condition that momenta of particles are controlled, which will be proved in Section \ref{sec:thm_main1}.
This control of particle momenta exists in the two space dimensions, which is the second result and 
the demonstration of this will be shown in Section \ref{sec:thm_main2}.

\section{Preliminaries and the main results}\label{sec:pre_main}
Here are a few notational conventions. $C$ denotes a positive constant which
changes from line to line and may depend only on the initial data.
Similarly $C(t)$ denotes a positive nondecreasing function of time.
Also we use the norms
\begin{align*}
\|f(t)\| &= \sup \{f(t, x, p) : (x, p) \in \mathbb{R}^2 \times \mathbb{R}^2 \},\\
\|\phi(t)\| &= \sup \{|\phi(t, x)| : x \in \mathbb{R}^2\},\\
\interleave \phi(t) \interleave &= \sup \{\|\phi(\tau)\| : 0 \leq \tau \leq t \}.
\end{align*}
We also denote
\begin{align*}
\|D\phi(t)\| &= \sup \{|\partial_t\phi(t, x)|,\, |\partial_{x_i}\phi(t, x)| : x \in \mathbb{R}^2, i=1, 2\}\\
\|D^2\phi(t)\| &= \sup \{|\partial^2_t\phi(t, x)|, \,|\partial_t\partial_{x_i}\phi(t, x)|,\, 
			   	  |\partial_{x_i}\partial_{x_j}\phi(t, x)|: x \in \mathbb{R}^2, \, i, j=1, 2\}.
\end{align*}

Let us recall the representations for the electric and magnetic fields, $E_k$ (with $k=1, 2$) and $B$, in the case
of the relativistic Vlasov-Maxwell system in two space dimensions (Theorem 1, in \cite{GSh3}).
\begin{Lemma}
\begin{align*}
E_k(t, x) &= \tilde{E}^0_k - 2\int^t_0 \int_{\yxtt} \int \frac{(et_k) f}{\sqttyx} \,dp \,dy \,d\tau\\
 	   	  & \qquad - 2\int^t_0 \int_{\yxtt} \int 
		  	\frac{(E_1 + \hat{p}_2B, E_2 - \hat{p}_1B)f}{\sqttyx} 
					   \cdot \nabla_p(es_k) \,dp \,dy \,d\tau\\
B(t, x) & = \tilde{B}^0 + 2\int^t_0 \int_{\yxtt} \int \frac{(bt) f}{\sqttyx} \,dp \,dy \,d\tau\\
	 	& \qquad + 2\int^t_0 \int_{\yxtt} \int 
		  	\frac{(E_1 + \hat{p}_2B, E_2 - \hat{p}_1B)f}{\sqttyx} 
					   \cdot \nabla_p(bs) \,dp \,dy \,d\tau
\end{align*}
where $\tilde{E}^0_k$ and $\tilde{B}^0$ are Cauchy data terms and kernels are given by
\begin{align*}
et_k &= \frac{\xi_k + \hat{p}_k}{(1+|p|^2)(\xiDp)^2},
&es_k &= \frac{\xi_k + \hat{p}_k}{\xiDp},\\
bt &= \frac{\xi_1 \hat{p}_2 - \xi_2 \hat{p}_1}{(1+|p|^2)(\xiDp)^2},
&bs &= \frac{\xi_1 \hat{p}_2 - \xi_2 \hat{p}_1}{\xiDp}
\end{align*}
Here $\xi = \frac{y-x}{t-\tau}$.
\end{Lemma}
The next two propositions show that the derivatives of $\phi$ satisfy similar representations.

\begin{Proposition}\label{Prop:phi_t}
\begin{align*}
\partial_t \phi(t, x) &= \partial_t \phi_{\rm hom} - 2 \int_{\yxt}\int \frac{f^{\rm in}(y, p)}{\sqp(\xiDp)\sqtyx}\,dp \,dy\\
& \quad +2 \int^t_0 \int_{\yxtt} \int \frac{a^{\phi_t}(\xi, p) f(\tau, y, p)}{(t-\tau)\sqttyx}\,dp \,dy \,d\tau\\
& \quad -2 \int^t_0 \int_{\yxtt} \int \frac{b^{\phi_t}(\xi, p) S(\phi) f(\tau, y, p)}{\sqttyx}\,dp \,dy \,d\tau\\
& \quad -2 \int^t_0 \int_{\yxtt} \int \frac{c^{\phi_t}(\xi, p)\cdot(\nabla_x \phi) f(\tau, y, p)}{\sqttyx}\,dp \,dy \,d\tau
\end{align*}
\begin{align*}
a^{\phi_t}(\xi, p) & = \frac{\hat{p}\cdot (\xi + \hat{p})}{\sqp(\xiDp)^2} = p \cdot (et_1, et_2)\\
b^{\phi_t}(\xi, p) & = \frac{1}{\sqp}\\
c^{\phi_t}(\xi, p) & = \frac{\xi + \hat{p}}{(1+|p|^2)^{3/2}(\xiDp)^2}
\end{align*}
\end{Proposition}
\begin{Proposition}\label{Prop:phi_x1}
\begin{align*}
\partial_{x_1} \phi(t, x) &= \partial_{x_1}\phi_{\rm hom}
			   -2 \int_{\yxt}\int \frac{\xi_1 f^{\rm in}(y, p)}{\sqp(\xiDp)\sqtyx}\,dp \,dy\\
& \quad -2 \int^t_0 \int_{\yxtt} \int \frac{a^{\phi_{x_1}}(\xi, p) f(\tau, y, p)}{(t-\tau)\sqttyx}\,dp \,dy \,d\tau\\
& \quad -2 \int^t_0 \int_{\yxtt} \int \frac{b^{\phi_{x_1}}(\xi, p) S(\phi) f(\tau, y, p)}{\sqttyx}\,dp \,dy \,d\tau\\
& \quad -2 \int^t_0 \int_{\yxtt} \int \frac{c^{\phi_{x_1}}(\xi, p) \cdot (\nabla_x \phi) f(\tau, y, p)}{\sqttyx}\,dp \,dy \,d\tau
\end{align*}
\begin{align*}
a^{\phi_{x_1}}(\xi, p) & = \frac{(\xi_1 +\hat{p}_1) - \hat{p}_2(\xi_1 \hat{p}_2 - \xi_2 \hat{p}_1 )}{\sqp(\xiDp)^2}
					   	 = \sqp[(et_1) - \hat{p}_2 (bt)]\\
b^{\phi_{x_1}}(\xi, p) & = \frac{\xi_1}{\sqp} = \xi_1 b^{\phi_t}(\xi, p)\\
c^{\phi_{x_1}}(\xi, p) & = \frac{\xi_1(\xi + \hat{p})}{(1+|p|^2)^{3/2}(\xiDp)^2} = \xi_1 c^{\phi_t}(\xi, p)
\end{align*}
\end{Proposition}
The representation for $\partial_{x_2}\phi$ is almost identical to the one for $\partial_{x_1}\phi$ and so we omit it.
The proof of Proposition \ref{Prop:phi_x1} is provided in the appendix.

One basic property of the Vlasov equation is that the distribution $f$ is constant along the characteristic.
However, this is no longer true in Nordstr\"om-Vlasov system. Nevertheless one can have a similar property.
The following lemma is from \cite{CaLe}. 
It is true also for the two space dimensions and the proof is shown in the appendix.
\begin{Lemma}\label{Lemm:f}
Let $f^{\rm in}\in C^1_b(\mathbb{R}^4)$, $\phi_{0}^{\rm in}\in C^3_b(\mathbb{R}^2)$
and $\phi_1^{\rm in}\in C^2_b(\mathbb{R}^2)$. Then
$$ 
\|f(t)\| \leq C e^{ct} 
$$ 
for all $t\in \mathbb{R}$.
\end{Lemma}

Here is the main result of this paper :
\begin{Theorem}\label{global}
Let $f^{\rm in}\in C^1_b(\mathbb{R}^4)$ with compact support in $p$, $\phi_{0}^{\rm in}\in C^3_b(\mathbb{R}^2)$
and $\phi_1^{\rm in}\in C^2_b(\mathbb{R}^2)$. Then
there exists a unique classical solution 
$(f,\phi)\in C^1([0, \infty)\times \mathbb{R}^4)\times C^2([0, \infty)\times \mathbb{R}^2)$ 
of the Nordstr\"om-Vlasov system (\ref{NV1})-(\ref{NV2}). 
\end{Theorem}

We will prove this main result by showing the following two theorems in the rest of the paper.
\begin{Theorem}\label{main_1}
Let $f^{\rm in}\in C^1_b(\mathbb{R}^4)$, $\phi^{\rm in}_0 \in C^3_b(\mathbb{R}^2)$ 
and $\phi^{\rm in}_1 \in C^2_b(\mathbb{R}^2)$.
Assume that there exists a nondecreasing function $C(t)$ for which
$$
f(t, x, p) = 0 \text{ if }|p| \geq C(t).
$$
Then there exists a unique classical solution 
$(f,\phi)\in C^1([0, \infty)\times \mathbb{R}^4)\times C^2([0, \infty)\times \mathbb{R}^2)$ 
of the Nordstr\"om-Vlasov system (\ref{NV1})-(\ref{NV2}). 
\end{Theorem}
\begin{Theorem}\label{main_2}
Assume the initial data from Theorem \ref{main_1}. Also we assume that $f^{\rm in}$ has compact support in $p$.
Then there exists a unique classical solution 
$(f,\phi)\in C^1([0, \infty)\times \mathbb{R}^4)\times C^2([0, \infty)\times \mathbb{R}^2)$ 
of the Nordstr\"om-Vlasov system (\ref{NV1})-(\ref{NV2}) satisfying 
$$
f(t, x, p) = 0 \text{ if } |p| \geq C(t)
$$
for some continuous function $C(t)$ and
$$
\|f(t)\| + \|\nabla_{(t, x, p)} f(t)\| + \|D \phi(t)\| + \|D^2 \phi(t)\| \leq C(t)
$$
for all $t \geq 0$.
\end{Theorem}
In the notation of the spaces of functions used above, the subscript  
$b$ means that all the derivatives up to the indicated order are bounded. 

\section{Proof of Theorem \ref{main_1}}\label{sec:thm_main1}
\subsection{Estimates on $D\phi$}\label{estimatesDphi}
\begin{Theorem}\label{Thm:phi_Dphi}
Assume that $f^{\rm in} \in C_b (\mathbb{R}^4)$, $\phi^{\rm in}_0 \in C^2_b (\mathbb{R}^2)$
and $\phi^{\rm in}_1 \in C^1_b (\mathbb{R}^2)$.
Assume that there exists a nondecreasing function $C(t)$ for which
$$
f(t, x, p) = 0  \text{ if } |p| \geq C(t).
$$
Then
$$\interleave \phi(t) \!\interleave + \interleave\! D\phi(t)\interleave \leq C(t).$$
\end{Theorem}
{\sc Proof : }  
The classical solution of (\ref{NV1}) is
\begin{equation}\label{eqn:phi}
\phi(t, x) = \phi_{\rm hom}(t, x) - 2 \int^t_0 \int_{\yxtt} \int \frac{f(\tau, y, p)}{\sqp\sqttyx}\,dp \,dy \,d\tau
\end{equation}
where
\begin{equation*}
\phi_{\rm hom} = \frac{1}{2\pi} \left[ \int_{\yxt} \frac{\phi^{\rm in}_1(y)\,dy}{\sqtyx}
	   				 + \frac{\partial}{\partial t}\Big(\int_{\yxt}\frac{\phi^{\rm in}_0(y) \,dy}{\sqtyx}\Big)\right]
\end{equation*}
is the solution of the homogeneous wave equation with data $\phi_0^{\rm in}$ and $\phi_1^{\rm in}$
and the second term in (\ref{eqn:phi}) is the solution of (\ref{NV1}) with trivial data.
Then with the assumption of data in the theorem, one can see that
\begin{equation}\label{eqn:esti.phi_hom}
\|\phi_{\rm hom}(t)\| \leq C(1+t)\big[ 
\|\phi^{\rm in}_0 \| + \|D \phi^{\rm in}_0 \|+\|\phi^{\rm in}_1\|\big] \leq C(1+t).
\end{equation}
With Lemma \ref{Lemm:f}, the second term in (\ref{eqn:phi}) becomes
$$
\int^t_0 \int_{\yxtt} \int_{\pC} \frac{f(\tau, y, p)\,dp \,dy \,d\tau}{\sqp\sqttyx} 
\leq C(t) \int^t_0 \|f(\tau)\| (t-\tau) \,d\tau \leq C(t).
$$
Therefore we have
$$
\interleave \phi(t) \interleave \leq C(t).
$$
Using the fact that $D\phi_{\rm hom}$ satisfies the homogeneous wave equation, we get 
$\|D \phi_{\rm hom}(t)\| \leq C(t).$
Note that given $|p|< C(t)$, we have $(\xiDp)^{-1} \leq C(t).$
Then one can also see that 
the second terms of the representations $\partial_t \phi$ and $\partial_{x_1} \phi$ are bounded by $C(t)$.
Also using a similar argument to the kernels in Propositions \ref{Prop:phi_t} and \ref{Prop:phi_x1}, we obtain
$$|f(\tau, y, p)|\big(|a^{\phi_t}|+|b^{\phi_t}|+|c^{\phi_t}|+|a^{\phi_{x_1}}|+|b^{\phi_{x_1}}|+|c^{\phi_{x_1}}|\big)
\leq C(\tau).$$
Therefore
\begin{align*}
\| D\phi(t) \| 
& \leq C(t) + C(t)\int^t_0 \int_{\yxtt}\int_{\pC} \frac{\big[(t-\tau)^{-1}+\|D\phi(t)\|\big]}{\sqttyx} \,dp \,dy \,d\tau\\
& \leq C(t) + C(t)\int^t_0 \big[1+\|D\phi(\tau)\|(t-\tau)\big] \,d\tau.
\end{align*}
By Gronwall's inequality,
$\|D\phi(t)\| \leq C(t)$. 
\hfill$\Box$

\subsection{Estimates on $D^2 \phi$}
\begin{Theorem}\label{Thm:phi_D2phi}
Let $(f, \phi)$ be as in Theorem \ref{Thm:phi_Dphi} and assume that
$f^{\rm in} \in C^1_b(\mathbb{R}^4)$, $\phi^{\rm in}_0 \in C^3_b(\mathbb{R}^2)$
and $\phi^{\rm in}_1 \in C^2_b(\mathbb{R}^2)$.
Then
$$
\interleave D^2\phi(t)\interleave \leq C(t) \big[1 + \ln^*(t \interleave\nabla_{(x, p)}f(t)\interleave)\big] 
$$
where 
\begin{equation*}
\ln^*(t) = \left\{ 
\begin{aligned}
&0 &&\text{if } 0 \leq t \leq 1, \\
&\ln(t) &&\text{if } 1 <t.
\end{aligned}
\right.
\end{equation*}
\end{Theorem}

{\sc Proof : } Here we will prove the estimate for $\partial_{x_1}^2 \phi$.
The other derivatives can be obtained with the same argument presented in the following.
First, in the representation of $\partial_{x_1} \phi$, define
\begin{equation*}
A^{\phi_{x_1}} := \int^t_0 \int_{\yxtt}\int \frac{a^{\phi_{x_1}}(\xi, p) f(\tau, y, p)}{(t-\tau)\sqttyx}\,dp \, dy \,d\tau.
\end{equation*}
Then using (\ref{def:x1}) we obtain
\begin{align*}
\partial_{x_1} A^{\phi_{x_1}} 
&= \int^t_0 \int_{\yxtt} \int 
   \frac{a^{\phi_{x_1}}(\xi, p) \partial_{x_1} f(\tau, y, p)}{(t-\tau)\sqttyx}\,dp \, dy \,d\tau\\
&= \int^t_0 \int_{\yxtt} \int
   \frac{a^{\phi_{x_1}}(\xi, p) \xi_1 Sf(\tau, y, p)}{(\xiDp)(t-\tau)\sqttyx}\,dp \, dy \,d\tau\\
&\quad + \int^t_0 \int_{\yxtt} \int
   \frac{a^{\phi_{x_1}}(\xi, p)\big[(1+\xi_2 \hat{p}_2)T_1 - \xi_1 \hat{p_2}T_2\big]f(\tau, y, p)}
   {(\xiDp)(t-\tau)^2}\,dp \, dy \,d\tau\\
&:= \partial_{x_1} A^{\phi_{x_1}}S + \partial_{x_1} A^{\phi_{x_1}}T.
\end{align*}
Also $\delta \in (0, t)$ and $\epsilon \in (0, 1)$.
Using (\ref{prop:Tk}), we get from $\partial_{x_1} A^{\phi_{x_1}}T$
\begin{align}
&\int^{t-\delta}_0 \int_{\yxtte} \int
   \frac{a^{\phi_{x_1}}(\xi, p)\big[(1+\xi_2 \hat{p}_2)T_1 - \xi_1 \hat{p_2}T_2\big]f(\tau, y, p)}
   {(\xiDp)(t-\tau)^2}\,dp \, dy \,d\tau \label{eqn:for2ndterm}\\
& \quad = - \int^{t-\delta}_0 \int_{\yxtte} \int
  		  \frac{f(\tau, y, p)}{\sqxi}\left[(-\xi_1, 1, 0) \cdot \nabla_{(\tau, y)}a^{\phi_{x_1}}_1(\xi, p)\right.\notag\\
&\qquad\qquad\qquad \left.-(-\xi_2, 0, 1) \cdot\nabla_{(\tau, y)}a^{\phi_{x_1}}_2(\xi, p)\right] \,dp \,dy \,d\tau \notag\\
& \qquad + \int^{t-\delta}_0 \int_{|y-x|=(1-\epsilon)(t-\tau)} \int 
  		 fA \cdot \left(1-\epsilon, \frac{y_1-x_1}{|y-x|}, \frac{y_2-x_2}{|y-x|}\right) \,dp \,dS_y \,d\tau \notag\\
& \qquad + \int_{|y-x|<(1-\epsilon)\delta} \left. \int fA \right|_{\tau = t-\delta} \cdot (1, 0, 0) \,dp \,dy
+ \int_{|y-x|<(1-\epsilon)t} \left. \int fA \right|_{\tau = 0} \cdot (-1, 0, 0) \,dp \,dy,\notag
\end{align}
where
$$
a^{\phi_{x_1}}_1(\xi, p) := \frac{(1+\xi_2 \hat{p}_2) a^{\phi_{x_1}}}{(\xiDp)(t-\tau)^2},\qquad
a^{\phi_{x_1}}_2(\xi, p) := \frac{\xi_1 \hat{p}_2 a^{\phi_{x_1}}}{(\xiDp)(t-\tau)^2}
$$
and
$$
A := [a^{\phi_{x_1}}_1(\xi, p)(-\xi_1, 1, 0) - a^{\phi_{x_1}}_2(\xi, p)(-\xi_2, 0, 1)](1-|\xi|^2)^{-1/2}.
$$
Applying to the second term in (\ref{eqn:for2ndterm}) the similar argument in (\ref{eqn:epsilon})
and then by letting $\epsilon \rightarrow 0^+$, we obtain
\begin{align}\label{eqn:forlastterm}
&\int^{t-\delta}_0 \int_{\yxtt} \int
   \frac{a^{\phi_{x_1}}(\xi, p)\big[(1+\xi_2 \hat{p}_2)T_1 - \xi_1 \hat{p_2}T_2\big]f(\tau, y, p)}
   {(\xiDp)(t-\tau)^2}\,dp \, dy \,d\tau\\
& \quad = - \int^{t-\delta}_0 \int_{\yxtt} \int
  		  \frac{f(\tau, y, p)}{\sqxi}\left[(-\xi_1, 1, 0) \cdot \nabla_{(\tau, y)}a^{\phi_{x_1}}_1(\xi, p)\right.\notag\\
& \qquad\qquad\qquad\left.-(-\xi_2, 0, 1) \cdot\nabla_{(\tau, y)}a^{\phi_{x_1}}_2(\xi, p)\right] \,dp \,dy \,d\tau\notag\\
& \qquad + \int_{|y-x|<\delta} \left. \int fA \right|_{\tau = t-\delta} \cdot (1, 0, 0) \,dp \,dy
+ \int_{|y-x|<t} \left. \int fA \right|_{\tau = 0} \cdot (-1, 0, 0) \,dp \,dy.\notag
\end{align}
Note that 
$$
\left|A \cdot (1, 0, 0)\right| \leq C(t)(t-\tau)^{-2}(1-|\xi|^2)^{-1/2}.
$$
So we obtain
\begin{equation*}
\left|\int_{|y-x|<\delta} \left. \int fA \right|_{\tau = t-\delta} \cdot (1, 0, 0) \,dp \,dy\right|
\leq C(t) \int_{|y-x|<\delta} \delta^{-1} (\delta^2-|y-x|^2)^{-1/2}\,dy 
\leq C(t)
\end{equation*}
and the same estimate holds for the last term in (\ref{eqn:forlastterm}).
Now we compute
\begin{align*}
&\left|(-\xi_1, 1, 0) \cdot \nabla_{(\tau, y)}a^{\phi_{x_1}}_1(\xi, p)
				-(-\xi_2, 0, 1) \cdot\nabla_{(\tau, y)}a^{\phi_{x_1}}_2(\xi, p)\right|\\
& \quad = \left|-\xi_1 \Big(\frac{2a^{\phi_{x_1}}_1(\xi, p)}{(t-\tau)^3} + 
			 \nabla_\xi a^{\phi_{x_1}}_1(\xi, p)\cdot \frac{\partial\xi}{\partial \tau}\Big)
			 + \nabla_\xi a^{\phi_{x_1}}_1(\xi, p)\cdot \frac{\partial\xi}{\partial y_1}\right.\\
& \qquad\quad \left.+ \xi_2 \Big(\frac{2a^{\phi_{x_1}}_2(\xi, p)}{(t-\tau)^3} + 
			 \nabla_\xi a^{\phi_{x_1}}_2(\xi, p)\cdot \frac{\partial\xi}{\partial \tau}\Big)
			 - \nabla_\xi a^{\phi_{x_1}}_2(\xi, p)\cdot \frac{\partial\xi}{\partial y_2}\right|
\leq \frac{C(t)}{(t-\tau)^3}.			 	
\end{align*}
So we get
\begin{align*}
&\left|\int^{t-\delta}_0 \int_{\yxtt} \int
  		  \frac{f(\tau, y, p)}{\sqxi}\left[(-\xi_1, 1, 0) \cdot \nabla_{(\tau, y)}a^{\phi_{x_1}}_1
-(-\xi_2, 0, 1) \cdot\nabla_{(\tau, y)}a^{\phi_{x_1}}_2
\right] \,dp \,dy \,d\tau\right|\\
& \qquad \leq \int^{t-\delta}_0 \int_{\yxtt} \frac{C(t)}{(t-\tau)^3\sqxi} \,dy \,d\tau = C(t) \ln\frac{t}{\delta}.
\end{align*}
Therefore (\ref{eqn:forlastterm}) becomes
\begin{equation}\label{eqn:bottom_A}
\int^{t-\delta}_0 \int_{\yxtt} \int
   \frac{a^{\phi_{x_1}}(\xi, p)\big[(1+\xi_2 \hat{p}_2)T_1 - \xi_1 \hat{p_2}T_2\big]f(\tau, y, p)}
   {(\xiDp)(t-\tau)^2}\,dp \, dy \,d\tau \leq C(t) \Big(1+ \ln\frac{t}{\delta}\Big).
\end{equation}
For the tip of the cone, with (\ref{NV2}) and Theorem \ref{Thm:phi_Dphi} note that
$$
\left|\frac{a^{\phi_{x_1}}(\xi, p)\big[(1+\xi_2 \hat{p}_2)T_1 - \xi_1 \hat{p_2}T_2\big]f(\tau, y, p)}
   {(\xiDp)} \right| \leq \frac{C(t)(1+ \interleave \nabla_{(x, p)} f(t) \interleave)}{\sqxi}.
$$
So we have
\begin{align}\label{eqn:top_A}
&\left|\int^t_{t-\delta} \int_{\yxtt} \int
   \frac{a^{\phi_{x_1}}(\xi, p)\big[(1+\xi_2 \hat{p}_2)T_1 - \xi_1 \hat{p_2}T_2\big]f(\tau, y, p)}
   {(\xiDp)(t-\tau)^2}\,dp \, dy \,d\tau \right|\\
& \qquad \leq C(t) (1+ \interleave \nabla_{(x, p)} f(t) \interleave) 
		   	  \int^t_{t-\delta} \int_{\yxtt} \int_{\pC} \frac{\,dp \,dy \,d\tau}{(t-\tau)^2 \sqxi}
= C(t) (1+ \delta \interleave \nabla_{(x, p)} f(t) \interleave ).\notag
\end{align}
Therefore collecting (\ref{eqn:bottom_A}) and (\ref{eqn:top_A}) we obtain
$$
\partial_{x_1} A^{\phi_{x_1}} T 
\leq C(t) \big[1 +\ln\frac{t}{\delta} + \delta \interleave \nabla_{(x, p)} f(t) \interleave \big]
$$
and taking 
$
\delta = \min \{ t, \interleave \nabla_{(x, p)} f(t) \interleave^{-1}\}
$
we get
\begin{equation}\label{eqn:dx1_AT}
\partial_{x_1} A^{\phi_{x_1}} T 
\leq C(t) \big[1 +\ln^*(t \interleave \nabla_{(x, p)} f(t) \interleave) \big].
\end{equation}
Recall (\ref{eq:Sf}) : 
\begin{equation}\label{eq2:Sf}
Sf = F(t, x, p) \cdot \nabla_p f + 3 (S \phi) f
\end{equation}
where $$ F(t, x, p) := (S \phi) p + \frac{\nabla_x \phi}{\sqp}.$$
Then
\begin{align*}
\partial_{x_1} A^{\phi_{x_1}} S 
&= \int^t_0 \int_{\yxtt} \int
   \frac{a^{\phi_{x_1}}(\xi, p) \xi_1 Sf(\tau, y, p)}{(\xiDp)(t-\tau)\sqttyx}\,dp \, dy \,d\tau\\
&= \int^t_0 \int_{\yxtt} \int
   \frac{a^{\phi_{x_1}}(\xi, p) \xi_1 F(\tau, y, p) \cdot \nabla_pf(\tau, y, p)}
   							  {(\xiDp)(t-\tau)\sqttyx}\,dp \, dy \,d\tau\\
&\quad + \int^t_0 \int_{\yxtt} \int
   \frac{3 a^{\phi_{x_1}}(\xi, p) S(\phi) \xi_1 f(\tau, y, p)}
   							  {(\xiDp)(t-\tau)\sqttyx}\,dp \, dy \,d\tau.
\end{align*}
Note that
$
\nabla_p F = 2 S(\phi).
$
\begin{align*}
\partial_{x_1} A^{\phi_{x_1}}S 
&= - \int^t_0 \int_{\yxtt} \int
   \nabla_p\left(\frac{a^{\phi_{x_1}}(\xi, p)}{\xiDp}\right) \cdot
   \frac{\xi_1 F(\tau, y, p) f(\tau, y, p)}{(t-\tau)\sqttyx}\,dp \,dy \,d\tau\\
&\quad + \int^t_0 \int_{\yxtt} \int 
	  \frac{a^{\phi_{x_1}}(\xi, p) S(\phi) \xi_1 f(\tau, y,p)}{(\xiDp)(t-\tau)\sqttyx}\,dp \, dy \,d\tau.
\end{align*}
By Theorem \ref{Thm:phi_Dphi}, one can see that
\begin{equation}\label{eqn:dx1_AS}
|\partial_{x_1} A^{\phi_{x_1}} S| \leq C(t).
\end{equation}
Now collecting (\ref{eqn:dx1_AT}) and (\ref{eqn:dx1_AS}) we get
\begin{equation}
|\partial_{x_1} A^{\phi_{x_1}}| \leq C(t)\big[1+\ln^*(t\interleave\nabla_{(x, p)}f(t)\interleave) \big].
\end{equation}
Define
$$
B^{\phi_{x_1}} := \int^t_0 \int_{\yxtt} \int \frac{b^{\phi_{x_1}}(\xi, p) S(\phi)f(\tau, y, p)}{\sqttyx} \,dp \,dy \,d\tau.
$$
Then
$$
\partial_{x_1} B^{\phi_{x_1}} 
= \int^t_0 \int_{\yxtt} \int \frac{b^{\phi_{x_1}}(\xi, p) \big[\partial_{x_1}[S(\phi)]f + S(\phi)\partial_{x_1} f\big]}
  		   				{\sqttyx} \,dp \,dy \,d\tau.
$$
The first term becomes :
\begin{equation}\label{eqn:1st}
\left| \int^t_0 \int_{\yxtt} \int \frac{b^{\phi_{x_1}}(\xi, p) \partial_{x_1}[S(\phi)]f}
	   			{\sqttyx} \,dp \,dy \,d\tau \right|
\leq C(t) \int^t_0  \| D^2 \phi(\tau) \| \,d\tau.
\end{equation}
Using (\ref{def:x1}) the second term becomes :
\begin{align*}
&\int^t_0 \int_{\yxtt} \int \frac{b^{\phi_{x_1}}(\xi, p) S(\phi) \partial_{x_1} f}{\sqttyx} \,dp \,dy \,d\tau\\
&\quad = \int^t_0 \int_{\yxtt} \int 
	   	 \frac{\xi_1 b^{\phi_{x_1}}(\xi, p) S(\phi) Sf}{(\xiDp)\sqttyx} \,dp \,dy \,d\tau\\
& \qquad + \int^t_0 \int_{\yxtt} \int
  		   \frac{b^{\phi_{x_1}}(\xi, p)S(\phi) \big[(1+\xi_2 \hat{p}_2)T_1 - \xi_1 \hat{p}_2 T_2\big]f}
		   	{(\xiDp)(t-\tau)} \,dp \,dy \,d\tau\\
& \quad := \partial_{x_1}B^{\phi_{x_1}}S + \partial_{x_1}B^{\phi_{x_1}}T.
\end{align*}
Again use (\ref{eq2:Sf}) for the term $\partial_{x_1}B^{\phi_{x_1}}S$ :
\begin{align*}
\partial_{x_1} B^{\phi_{x_1}}S 
&= - \int^t_0 \int_{\yxtt} \int
   	 \nabla_p\left(\frac{b^{\phi_{x_1}}(\xi, p)}{\xiDp}\right) \cdot
	 \frac{\xi_1 F(\tau, y, p)S(\phi)f}{\sqttyx}\,dp \,dy \,d\tau\\
&\quad + \int^t_0 \int_{\yxtt}\int \frac{\xi_1 b^{\phi_{x_1}}(\xi, p) \big[S(\phi)\big]^2 f}
	   	 {(\xiDp)\sqttyx}\,dp \,dy \,d\tau.
\end{align*}
So one can see that
\begin{equation}\label{eqn:BS}
|\partial_{x_1} B^{\phi_{x_1}}S| \leq C(t).
\end{equation}
For the term $\partial_{x_1}B^{\phi_{x_1}}T$, let us carry out the computation with the term involved $T_1$. 
The argument is same for the term with $T_2$ and so will be omitted.
Let 
$$
B:= \frac{b^{\phi_{x_1}}(\xi, p)(1+\xi_2 \hat{p}_2)(-\xi_1, 1, 0)}
	{(\xiDp)(t-\tau)\sqxi}.
$$
Note that for $\epsilon \in (0, 1)$ we have
\begin{align*}
&\int^t_0 \int_{\yxtte} \int \frac{b^{\phi_{x_1}}(\xi, p) S(\phi)(1+\xi_2 \hat{p}_2)T_1 f}
		 			   		{(\xiDp)(t-\tau)}\,dp \,dy \,d\tau\\
&\quad = -\int^t_0 \int_{\yxtte} \int
   			 			   \nabla_{(\tau, y)}\left[\frac{b^{\phi_{x_1}}(\xi, p)S(\phi)(1+\xi_2 \hat{p}_2)}
						   {(\xiDp)(t-\tau)}\right] \cdot \frac{(-\xi_1, 1, 0)f}{\sqxi} \,dp \,dy \,d\tau\\
&\qquad + \int^t_0 \int_{|y-x|=(1-\epsilon)(t-\tau)} \int
	   	 S(\phi) fB \cdot \left(1-\epsilon, \frac{y_1 - x_1}{|y-x|}, \frac{y_2 - x_2}{|y-x|}\right)\,dp \,dS_y \,d\tau\\
&\qquad + \int^t_0 \int_{|y-x|<(1-\epsilon)t} \int\left.
	   	 S(\phi) fB \right|_{\tau=0} \cdot (-1, 0, 0)\,dp \,dy.							
\end{align*}
Applying the similar argument in (\ref{eqn:epsilon}) to the second term
and then by letting $\epsilon \rightarrow 0^+$, we obtain
\begin{align*}
&\int^t_0 \int_{\yxtt} \int \frac{b^{\phi_{x_1}}(\xi, p) S(\phi)(1+\xi_2 \hat{p}_2)T_1 f}
		 			   		{(\xiDp)(t-\tau)}\,dp \,dy \,d\tau\\
&\quad = -\int^t_0 \int_{\yxtt} \int
   			 			   \nabla_{(\tau, y)}\left[\frac{b^{\phi_{x_1}}(\xi, p)S(\phi)(1+\xi_2 \hat{p}_2)}
						   {(\xiDp)(t-\tau)}\right] \cdot \frac{(-\xi_1, 1, 0)f}{\sqxi} \,dp \,dy \,d\tau\\
&\quad + \int^t_0 \int_{\yxt} \int\left.
	   	 S(\phi) fB \right|_{\tau=0} \cdot (-1, 0, 0)\,dp \,dy.
\end{align*}
Then using Theorem \ref{Thm:phi_Dphi}, one can see that the last term is bounded by $C(t)$.
Let
$$
\tilde{b}^{\phi_{x_1}(\xi, p)} := b^{\phi_{x_1}}(\xi, p)(1+\xi_2 \hat{p}_2)(\xiDp)^{-1}.
$$
Compute the following :
\begin{align*}
&\left| \nabla_{(\tau, y)} \left[\frac{\tilde{b}^{\phi_{x_1}}(\xi, p)S(\phi)}{(t-\tau)}\right]
 \cdot (-\xi_1, 1, 0)\right|\\
& \quad \leq \left| -\xi_1 \left(\frac{\tilde{b}^{\phi_{x_1}}(\xi, p)}{(t-\tau)^2} 
			 			   + \frac{\nabla_\xi \tilde{b}^{\phi_{x_1}}(\xi, p) S(\phi)}{(t-\tau)} \cdot \frac{\partial \xi}{\partial \tau}
						   + \frac{\tilde{b}^{\phi_{x_1}}(\xi, p)(|\partial^2_t \phi|+|\partial_t \partial_{x_i}\phi|)}{(t-\tau)}\right)\right.\\
&\qquad\qquad\qquad \left. + \frac{\nabla_{\xi} \tilde{b}^{\phi_{x_1}}(\xi, p)S(\phi)}{(t-\tau)}\cdot \frac{\partial \xi}{\partial y_1}  
			  		+ \frac{\tilde{b}^{\phi_{x_1}}(\xi, p)}{(t-\tau)} \big(|\partial_{x_1}\partial_t \phi|+|\partial_{x_1} \partial_{x_i}\phi|\big)\right|\\
& \quad \leq C(t)\big[ (t-\tau)^{-2} + (t-\tau)^{-1} \|D^2\phi(\tau)\|\big].
\end{align*}
So we get
\begin{align}\label{eqn:BT}
&\left|\int^t_0 \int_{\yxtt} \int \frac{b^{\phi_{x_1}}(\xi, p) S(\phi)(1+\xi_2 \hat{p}_2)T_1 f}
		 			   		{(\xiDp)(t-\tau)}\,dp \,dy \,d\tau\right|\\
& \quad \leq C(t) + C(t) \int^t_0 \int_{\yxtt} \frac{(t-\tau)^{-1} + \|D^2 \phi(\tau)\|}{\sqttyx} \,dy \,d\tau\notag\\
& \quad = C(t) + C(t) \int^t_0 \big(1+(t-\tau)\|D^2\phi(\tau)\|\big)\,d\tau 
\leq C(t)\left[1 + \int^t_0 \|D^2\phi(\tau)\| \,d\tau \right].\notag
\end{align}
Therefore collecting (\ref{eqn:1st}), (\ref{eqn:BS}) and (\ref{eqn:BT}) we obtain
$$
|\partial_{x_1} B^{\phi_{x_1}}| \leq C(t)\left[1 + \int^t_0 \|D^2\phi(\tau)\|\,d\tau \right].
$$
A similar argument is applied to the $x_1$ derivative of the term involved with the kernel $c^{\phi_{x_1}}$ 
in the representation of $\phi_{x_1}$.
Note that $\partial_{x_1}^2 \phi_{\rm hom}$ satisfies a homogeneous wave equation with the given initial data
and so it is bounded by $C(t)$. Since $\partial_{x_1} f^{\rm in}$ is bounded,
the $x_1$ derivative of the second term in the representation of $\phi_{x_1}$ is bounded by $C(t)$ as well.  
Therefore we obtain
$$
\|D^2\phi(t)\| \leq C(t)\left[1+\ln^*(t\interleave\nabla_{(x, p)}f(t)\interleave) + \int^t_0 \|D^2\phi(\tau)\|\,d\tau\right].
$$
So by Gronwall's inequality the theorem follows. \hfill $\Box$

\subsection{Estimates on $Df$ and Proof of Theorem \ref{main_1}}
\begin{Theorem}\label{thm:Df+D2phi}
Let $(f, \phi)$ and initial data be as in Theorem \ref{Thm:phi_D2phi}.
Then
$$
\|\nabla_{(t, x, p)}f(t)\|+\|D^2\phi(t)\|\leq C(t).
$$
\end{Theorem} 
{\sc Proof :} First we assume more smoothness on the initial data, i.e., $f^{\rm in} \in C^2$,
$\phi^{\rm in}_0 \in C^4$ and $\phi^{\rm in}_1 \in C^3$. 
Then applying $\partial_{x_1}$ to the Vlasov equation (\ref{NV2}) and integrating it along 
the characteristics, by Lemma \ref{Lemm:f} and Theorem \ref{Thm:phi_Dphi} we get
$$
|\partial_{x_1}f(t, x, p)| \leq \|\partial_{x_1}f^{\rm in}\| + C(t)\int^t_0 \big(1+\|D^2 \phi(\tau)\|\big)
					\big(1+\|\nabla_{(x, p)}f(\tau)\|\big) \,d\tau.
$$ 
Similarly for $\partial_{p_1} f$ we have
$$
|\partial_{p_1}f(t, x, p)| \leq \|\partial_{p_1}f^{\rm in}\| + C(t)\int^t_0 (1+\|\nabla_{(x, p)}f(\tau)\|) \,d\tau.
$$ 
Therefore using Theorem \ref{Thm:phi_D2phi}, we get
$$
\|\nabla_{(x, p)}f(t)\| \leq \|\nabla_{(x, p)}f^{\rm in}\| + C(t)\int^t_0 
\big[1 + \ln^*(\tau \interleave\nabla_{(x, p)}f(\tau)\interleave)\big]
\big(1+\interleave\nabla_{(x, p)}f(\tau)\interleave\big) \,d\tau.
$$
This equation is still satisfied with our original data in the theorem by a limiting argument.
For fixed $t$ and $s\in [0, t]$, consider
\begin{equation}\label{eqn:nabla_f}
3+\interleave\nabla_{(x, p)}f(s)\interleave \leq Q(s)
\end{equation}
where $Q$ is defined by
$$
Q(s) := (3+\|\nabla_{(x, p)}f^{\rm in}\|)+ C(t)\int^s_0 Q(\tau)\ln Q(\tau) \,d\tau.
$$
Then one can see that
$$
Q(s) = \exp \big(e^{C(t)s} \ln(3+\|\nabla_{(x, p)}f^{\rm in}\|)\big).
$$
So taking $s=t$ with (\ref{eqn:nabla_f}) we obtain
\begin{equation}\label{esti:f_xp}
\interleave \nabla_{(x, p)}f(t)\interleave \leq C(t).
\end{equation}
The bound on $\|\partial_t f(t)\|$ follows by (\ref{NV2}), (\ref{esti:f_xp}), Lemma \ref{Lemm:f} and Theorem \ref{Thm:phi_Dphi} : 
$$
\|\partial_t f(t)\| \leq \|\nabla_x f(t)\| + \|D\phi(t)\| \,\|\nabla_p f(t)\| +\|D\phi(t)\| \,\|f(t)\| \leq C(t).
$$
With Theorem \ref{Thm:phi_D2phi}, the proof of the theorem completes. \hfill $\Box$

{\sc Proof of Theorem \ref{main_1}} : In \cite{CaRe}, 
the three dimensional version of the iteration scheme is presented and 
the convergence is shown. Even though the representations of the derivatives of $\phi$ are different in
the two dimensional case, the iteration scheme and the proof of the convergence in \cite{CaRe}
can be applied directly with Lemma \ref{Lemm:f}, Theorems \ref{Thm:phi_Dphi} and \ref{thm:Df+D2phi},
to end the argument of the existence of solution. 
We also refer to the same reference for the uniqueness of the solution. \hfill $\Box$

\section{Proof of Theorem \ref{main_2}}\label{sec:thm_main2}
In the previous section, it was shown that the solution is continued as long as the $p$ support of $f$
remains bounded for bounded time. By the assumption that $f^{\rm in}$ has compact support in $p$,
there is a smooth solution with
$$
f(t, x, p) = 0 \text{ if } |p| \geq C(t),
$$
on some time interval $[0, T)$. Without loss of generality 
we take $T$ to be maximal and consider $t \in [0, T)$ for the rest of the paper. Now define
$$
P(t) = \sup\{|p|:f(s, x, p) \neq 0 \text{ for some } (s, x) \in [0, t] \times \mathbb{R}^2\} + 3.
$$
Note that $ P(t) \leq C(t)$ implies that $T=\infty$ and so the bounds stated in 
Theorems \ref{Thm:phi_Dphi} and \ref{thm:Df+D2phi} hold for all $t$. 
Therefore once we achieve that $P(t) \leq C(t)$ then the theorem follows.

\subsection{Preliminaries and Lemmas}
We present some lemmas and notations to use frequently to prove Theorem \ref{main_2}.
To keep notations not to heavy we write
$ v \wedge w = v_1 w_2 - v_2 w_1$, for any two vectors $(v_1, v_2)$ and $(w_1, w_2)$. 
and also define $\omega = (y-x)/|y-x|.$

\begin{Lemma}\label{prop:1+p2}
$$(1+|p|^2)^{-1} \leq 2 (\xiDp).$$
\end{Lemma}
{\sc Proof :} The lemma follows by the fact that 
$$\sqp (\xiDp) = \sqp + \xi\cdot p
= \frac{1+|p|^2 - (\xi\cdot p)^2}{\sqp - \xi\cdot p} \geq 1/(2 \sqp).
$$ \hfill $\Box$

The following lemma is from \cite{GSh4}. We state it without the proof.
\begin{Lemma}\label{prop:pw}
$$(\hat{p}\wedge\omega)^2 \leq 2(\xiDp).$$
\end{Lemma}

Now for the next lemma, we define the energy density $e$ by  
$$
e(t, x) := 4\pi \int \sqp f(t, x, p) \,dp + \frac{1}{2} (\partial_t \phi)^2 + \frac{1}{2} |\nabla_x \phi|^2.
$$
\begin{Lemma}
Let the assumptions of Theorem \ref{main_2} hold. Then for each $R \geq 0$,
\begin{align}
& \sup_{x \in \mathbb{R}^2} \int_{|y-x|<R} e(t, y) \,dy \leq C(R+t)^2, \label{lemma1}\\
& \sup_{x \in \mathbb{R}^2} \int^t_0 \int_{|y-x|=t-\tau+R} 
  \left(\frac{1}{2}(\omega \wedge \nabla_x \phi)^2+
  \frac{1}{2}(\partial_t \phi - \nabla_x\phi\cdot\omega)^2 \notag \right.\\
& \qquad\qquad \left. +  4\pi \int \sqp (1+\hat{p}\cdot \omega) \,dp \right) \,dS_y \,d\tau \leq C(R+t)^2, \label{lemma2}\\
& \sup_{x \in \mathbb{R}^2} \int_{|y-x|<R} \left( \int \frac{f(t, x, p)}{\sqp} \,dp\right)^3 \,dy \leq C(t)(R+t)^2.\label{lemma4}
\end{align}
\end{Lemma}
{\sc Proof :} 
First note that
\begin{equation}\label{eqn:forEid}
\int_{|y-x| \leq R+t} e(0, y) \,dy \leq C(R+t)^2.
\end{equation}
We have the energy identity :
$$
\partial_t e(t, x) + \nabla_x \cdot \left(-\partial_t \phi \nabla_x\phi
		   		+ 4\pi \int pf(t, x, p) \,dp \right) =0.
$$
So we have
\begin{align*}
0 &= \int^t_0 \int_{|y-x| < t-\tau + R} \left[\partial_\tau e + \nabla_y \cdot
  	 		  \Big(-\partial_t \phi \nabla_x \phi + 4 \pi \int p f(\tau, y, p)\,dp \Big)\right]\,dy \,d\tau\\
  &= \int^t_0 \int_{|y-x|=t-\tau+R} \left[e+\omega\cdot\Big(-\partial_t \phi \nabla_x \phi 
  		   +4 \pi \int p f(\tau, y, p) \,dp\Big)\right] \,dS_y \,d\tau\\
  & \quad - \int_{|y-x|<t+R} e(0, y)\,dy + \int_{|y-x|<R} e(t, y)\,dy.
\end{align*}
Also one can see that
\begin{align*}
&e+\omega \cdot \Big(- \partial_t \phi \nabla_x \phi + 4\pi \int p f(t, x, p) \, dp\Big)\\
&\qquad = \frac{1}{2}(\omega \wedge \nabla_x \phi)^2 + \frac{1}{2} (\partial_t \phi-\nabla_x \phi \cdot \omega)^2 
  + 4 \pi \int \sqp (1+\hat{p}\cdot \omega)f\,dp \geq 0.
\end{align*}
Therefore with (\ref{eqn:forEid}), we obtain (\ref{lemma1}) and (\ref{lemma2}).
Note that for each $r>0$,
$$
\int \frac{f(t, x, p)}{\sqp} \,dp \leq C(t) \int_{|p|<r} |p|^{-1}\,dp + r^{-2} \int_{|p|>r} \sqp f \,dp 
\leq C(t)(r+r^{-2}e).
$$
and so taking $r= e^{1/3}$ and with (\ref{lemma1}) we obtain (\ref{lemma4}). \hfill $\Box$

The following lemma is almost identical to Lemma 3 in \cite{GSh4},
except that we have Lemma \ref{Lemm:f}. So we state it without the proof.
\begin{Lemma}\label{lemma:sigmaS}
For $|\xi| < 1$, define
$$
\sigma_{BC} (t, x, \xi) : = \int \frac{f(t, x, p)}{\sqp (\xiDp)} \,dp.
$$
Then
$$ 0 \leq \sigma_{BC} \leq C(t)P(t)\min\{P(t), (1-|\xi|^2)^{-1/2}\}.$$
\end{Lemma}

\subsection{Fields estimates}
Let  $A^{\phi_l}$, $B^{\phi_l}$ and $C^{\phi_l}$ be terms with kernels 
$a^{\phi_l}$, $b^{\phi_l}$ and $c^{\phi_l}$ respectively
in the representations of the derivatives of $\phi$, where $l= t, x_1$ and $x_2$.  
\begin{Lemma}\label{lemma:fields}
$$
|B^{\phi_t}|+ |C^{\phi_t}| + |B^{\phi_{x_i}}|+|C^{\phi_{x_i}}| \leq C \int^t_0 \int_{\yxtt}\int
\frac{f\big(|\partial_t\phi|+|\nabla_x \phi| + (\xiDp)^{-1}|\omega \wedge \nabla_x \phi|\big)\,dp \,dy \,d\tau}{\sqp \sqttyx}
$$
where $i=1$ and $2$.
\end{Lemma}
{\sc Proof : } 
Define $\omega^\perp = (-\omega_2, \omega_1)$
and then for every $z \in \mathbb{R}^2$ we have
$$
z=(\omega \cdot z)\omega + ( \omega \wedge z) \omega^\perp, 
\quad z \cdot\omega^\perp = \omega\wedge z, 
\quad z \wedge\omega^\perp = \omega \cdot z.
$$
For fixed $\xi$ and $\hat{p}$ define $\mathcal{F} : \mathbb{R}\times\mathbb{R}^2 \rightarrow \mathbb{R}$ by
$$
\mathcal{F}(g, \vec{h}) := 
\xi_i(\xiDp)^2 (g+\hat{p} \cdot \vec{h}) + \xi_i (\xi+\hat{p})\cdot \vec{h}(1+|p|^2)^{-1}.
$$
Then from $b^{\phi_{x_i}}$ and $c^{\phi_{x_i}}$, we have
$$
b^{\phi_{x_i}} S(\phi) + c^{\phi_{x_i}}\cdot\nabla_x\phi 
				= \frac{\mathcal{F}(\partial_t \phi, \nabla_x\phi)}{\sqp(\xiDp)^2}.
$$
Using $\{(0, \omega^\perp), (1, \omega), (-1, \omega)\}$ as an orthogonal basis of $\mathbb{R}^3$ one can write
$$
\mathcal{F}(\partial_t\phi, \nabla_x \phi)= 
\big( A_1(0, \omega^\perp) + A_2(1, \omega) + A_3(-1, \omega) \big) \cdot (\partial_t\phi, \nabla_x \phi)
$$
for all $\partial_t\phi \in \mathbb{R}$ and $\nabla_x \phi \in \mathbb{R}^2$ where
$$
A_1=\mathcal{F}(0, \omega^\perp), \quad 2A_2=\mathcal{F}(1, \omega), \quad 2A_3=\mathcal{F}(-1, \omega).
$$
Now we estimate $A_1$, $A_2$ and $A_3$. First note that applying Lemmas  \ref{prop:pw} and \ref{prop:1+p2}  
to the first and the second terms respectively we have
\begin{equation}\label{esti:A1}
|A_1| = |\mathcal{F}(0, \omega^\perp)|
	  = |\xi_i| (\xiDp)^2 |\omega\wedge\hat{p}| + |\xi_i||\omega\wedge\hat{p}|(1+|p|^2)^{-1}
	  \leq C\big[(\xiDp)^2 + (\xiDp)\big].
\end{equation}
Also we get
\begin{align}
|2A_2| &= |\mathcal{F}(1, \omega)|
	  = |\xi_i|(\xiDp)^2|1+\hat{p}\cdot\omega|+|\xi_i||(\xi+\hat{p})\cdot\omega|(1+|p|^2)^{-1}\notag\\
	  &\leq C(\xiDp)^2+C(\xiDp)\big| |\xi| + \hat{p}\cdot\omega \big|\notag\\
	  &\leq C(\xiDp)^2 + C(\xiDp)\big| |\xi|-1 + (\xiDp) + (\omega-\xi)\cdot\hat{p} \big|\notag\\
	  &\leq C(\xiDp)^2 + C(\xiDp)\big( 1-|\xi| + (\xiDp) + |\omega-\xi|\big)
	  \leq C(\xiDp)^2 \label{esti:A2}
\end{align}
by the fact that
$
|\omega-\xi| = 1-|\xi| \leq \xiDp.
$
Similarly
\begin{equation}\label{esti:A3}
|2A_3| = |\mathcal{F}(-1, \omega)|
	  = |\xi_i|(\xiDp)^2|-1+\hat{p}\cdot\omega|+ |\xi_i||(\xi+\hat{p})\cdot\omega|(1+|p|^2)^{-1} \leq C(\xiDp)^2.
\end{equation}
Collecting these bounds (\ref{esti:A1}) - (\ref{esti:A3}), we have
\begin{align*}
|\mathcal{F}(\partial_t\phi, \nabla_x \phi)| 
&\leq C\big[(\xiDp)^2 + (\xiDp)\big]|(0, \omega^\perp)\cdot(\partial_t\phi, \nabla_x \phi)|\\
& \quad + C(\xiDp)^2|(1, \omega)\cdot(\partial_t\phi, \nabla_x \phi)| 
  		+ C(\xiDp)^2|(-1, \omega)\cdot(\partial_t\phi, \nabla_x \phi)|\\
&\leq C (\xiDp)^2 \big(|\partial_t\phi|+|\nabla_x \phi|\big) + C(\xiDp)|\omega \wedge \nabla_x \phi|.
\end{align*}
Therefore we obtain
$$
|B^{\phi_{x_i}}|+ |C^{\phi_{x_i}}|\leq C \int^t_0 \int_{\yxtt}\int
\frac{f\big(|\partial_t\phi|+|\nabla_x \phi| + (\xiDp)^{-1}|\omega \wedge \nabla_x \phi|\big) \,dp \,dy \,d\tau}{\sqp \sqttyx}.
$$
A same argument works for $B^{\phi_t}$ and $C^{\phi_t}$ and so the lemma follows. \hfill $\Box$

\begin{Lemma}\label{lemma:intSigmaS}
$$
\int^t_0 \int_{\yxtt} \frac{\sigma_{BC}(\tau, y, \xi) |\omega \wedge \nabla_x \phi(\tau, y)|\,dy \,d\tau}{\sqttyx} 
\leq C(t)P(t)\ln P(t).
$$
\end{Lemma}
{\sc Proof :} 
Let $r=|y-x|$ and  $s = (t-\tau-r)/2$ in the $r$ integration. 
We invert the order of the $\tau$ and $s$ and then change back to $r$ in the $\tau$ integration : 
\begin{align*}
& \int^t_0 \int_{\yxtt} \frac{\sigma_{BC}(\tau, y, \xi)|\omega \wedge \nabla_x \phi(\tau, y)|\,dy \,d\tau}{\sqttyx}\\
&\quad = \int^t_0 \int^{t-\tau}_0 \int_{|y-x|=r} 
  		\frac{\sigma_{BC}(\tau, y, \xi) |\omega \wedge \nabla_x \phi(\tau, y)|\,dS_y \,dr \,d\tau}{\sqrt{(t-\tau)^2-r^2}} \\
&\quad= \int^t_0 \int^{\frac{1}{2}(t-\tau)}_0 \int_{|y-x|=t-\tau-2s}
  		  \frac{\sigma_{BC}(\tau, y, \xi)|\omega \wedge \nabla_x \phi(\tau, y)|\,dS_y \,ds \,d\tau}{\sqrt{s} \sqrt{t-\tau-s}}\\	
&\quad = \int^{t/2}_0 \int^{t-2s}_0 \int_{|y-x|=t-\tau-2s}
  		  \frac{\sigma_{BC}(\tau, y, \xi)|\omega \wedge \nabla_x \phi(\tau, y)|\,dS_y \,d\tau \,ds}{\sqrt{s}\sqrt{t-\tau-s}}\\
&\quad = \int^{t/2}_0 \int^{t-2s}_0 \int_{|y-x|=r}
  		  \frac{\sigma_{BC}(t-r-2s, y, (y-x)(r+2s)^{-1})|\omega \wedge \nabla_x \phi(t-r-2s, y)| \,dS_y \,dr \,ds}
		  {\sqrt{s}\sqrt{r+s}}.
\end{align*}
Let $\epsilon \in (0, t/2]$ and consider $\tau \in (\epsilon, t/2)$. From Lemma \ref{lemma:sigmaS}
and with $|y-x|=r=t-\tau-2s$, we have
$$
\sigma_{BC}(t-r-2s, y, (y-x)(r+2s)^{-1})	\leq \frac{C(t)P(t)(r+2s)}{\sqrt{s}\sqrt{r+s}} 
					   						\leq C(t)P(t)\sqrt{r+s}/{\sqrt{s}}.
$$
Hence we get
\begin{align}\label{eq:temp_e}
&\int^{t/2}_\epsilon \int^{t-2s}_0 \int_{|y-x|=r}
	\frac{\sigma_{BC}(t-r-2s, y, (y-x)(r+2s)^{-1})|\omega \wedge \nabla_x \phi(t-r-2s, y)| \,dS_y \,dr \,ds}
	{\sqrt{s}\sqrt{r+s}}\\
& \quad \leq C(t)P(t) \int^{t/2}_\epsilon \int^{t-2s}_0 \int_{|y-x|=r}
  			 s^{-1}|\omega \wedge \nabla_x \phi(t-r-2s, y)| \,dS_y \,dr \,ds. \notag
\end{align}
By (\ref{lemma2}) and letting $r=t-\tau-2s$ we have
\begin{align}\label{eq:intepsilon}
&\int^{t-2s}_0 \int_{|y-x|=r} |\omega \wedge \nabla_x \phi(t-r-2s, y)|^2 \,dS_y \,dr\\
&\quad = \int^{t-2s}_0 \int_{|y-x|=t-\tau-2s} |\omega \wedge \nabla_x \phi(\tau, y)|^2 \,dS_y \,d\tau
\leq C(t-2s)^2 \leq Ct^2\notag.
\end{align}
So by Schwarz's inequality (\ref{eq:temp_e}) becomes
\begin{align}\label{eqn:Sigma+F1}
&\int^{t/2}_\epsilon \int^{t-2s}_0 \int_{|y-x|=r}
					 \frac{\sigma_{BC}(t-r-2s, y, (y-x)(r+2s)^{-1})|\omega \wedge \nabla_x \phi(t-r-2s, y)| \,dS_y \,dr \,ds}
					 {\sqrt{s}\sqrt{r+s}}\\
& \quad \leq C(t)P(t) \int^{t/2}_\epsilon
  			 	   \left(\int^{t-2s}_0 \int_{|y-x|=r} s^{-2} \,dS_y \,dr\right)^{1/2}\,ds
= C(t)P(t) \ln\frac{t}{2\epsilon}.\notag
\end{align}
Consider $\tau \in (0, \epsilon)$. From Lemma \ref{lemma:sigmaS},
$\sigma_{BC} \leq C(t)P^2(t)$ and by (\ref{eq:intepsilon}) we get
\begin{align}\label{eqn:Sigma+F2}
&\int^\epsilon_0 \int^{t-2s}_0 \int_{|y-x|=r}
					 \frac{\sigma_{BC}(t-r-2s, y, (y-x)(r+2s)^{-1})|\omega \wedge \nabla_x \phi(t-r-2s, y)| \,dS_y \,dr \,ds}
					 {\sqrt{s}\sqrt{r+s}}\\
& \;\leq C(t)P^2(t) \int^\epsilon_0 \int^{t-2s}_0 \int_{|y-x|=r}
  					|\omega \wedge \nabla_x \phi(t-r-2s, y)| [s(r+s)]^{-1/2} \,dS_y \,dr \,ds\notag\\
& \;\leq C(t)P^2(t) \int^\epsilon_0  \left(\int^{t-2s}_0 \int_{|y-x|=r}[s(r+s)]^{-1}\,dS_y \,dr\right)^{1/2} \,ds\notag
\leq C(t)P^2(t) \int^\epsilon_0 s^{-1/2} \,ds = C(t)P^2(t) \sqrt{\epsilon}.\notag
\end{align}
Collecting (\ref{eqn:Sigma+F1}) and (\ref{eqn:Sigma+F2}) we obtain that
$$
\int^t_0 \int_{\yxtt} \frac{\sigma_{BC}(\tau, y, \xi) |\omega \wedge \nabla_x \phi(\tau, y)|\,dy \,d\tau}{\sqttyx} 
 \leq C(t)P(t)\left( \ln\frac{t}{2\epsilon} + P(t)\sqrt{\epsilon} \right).
$$
Taking $\epsilon = \min \{t/2, P^{-2}(t)\}$ completes the proof. \hfill $\Box$
 
\begin{Proposition}\label{eqn:BC}
\begin{equation*}
|B^{\phi_t}|+ |C^{\phi_t}| + |B^{\phi_{x_i}}|+|C^{\phi_{x_i}}|
\leq C(t)P(t) \ln P(t) + C(t) \int^t_0 (\|\partial_t \phi(\tau)\|+\|\nabla_x \phi(\tau)\|) \,d\tau.
\end{equation*}
\end{Proposition}
{\sc Proof } : By Lemmas \ref{lemma:fields} and \ref{lemma:intSigmaS} we obtain
\begin{align*}
&|B^{\phi_t}|+ |C^{\phi_t}| + |B^{\phi_{x_i}}|+|C^{\phi_{x_i}}|\\
& \quad \leq C \int^t_0 \int_{\yxtt} \int 
  			   \frac{(|\partial_t \phi|+|\nabla_x \phi|) f \,dp \,dy \,d\tau}{\sqp \sqttyx}
			   + C \int^t_0 \int_{\yxtt} \frac{\sigma_{BC} |\omega \wedge \nabla_x \phi| \,dy \,d\tau}{\sqttyx}\\
&\quad \leq  C\int^t_0 (\|\partial_t\phi(\tau)\|+\|\nabla_x\phi(\tau)\|)
	   	 	      \int_{\yxtt}\int \frac{f \,dp\,dy\,d\tau}{\sqp\sqttyx} + C(t)P(t)\ln P(t).
\end{align*}
By (\ref{lemma4}) and H\"older's inequality , note that
\begin{align*}
&\int_{\yxtt}\int \frac{f(\tau, y, p) \,dp \,dy}{\sqp\sqttyx}\\
&\quad \leq \left(\int_{\yxtt}\left(\int \frac{f(\tau, y, p)}{\sqp}\,dp\right)^3 \,dy\right)^{1/3}
	   		\left(\int_{\yxtt} \big((t-\tau)^2-|y-x|^2\big)^{-3/4} \,dy\right)^{2/3} \leq C(t).
\end{align*}
Therefore the proposition follows. \hfill $\Box$

\begin{Lemma}\label{lemma:a_phi}
$$\int (|a^{\phi_t}|+|a^{\phi_{x_i}}|)f \,dp \leq C(t) \min \{P^3(t), P^{3/2}(t)e^{1/2}(t,x) (1-|\xi|^2)^{-1/4}\}.$$
\end{Lemma}
The proof of this lemma is almost identical to Lemma 5 in \cite{GSh4},
except the fact that the kernels $a^{\phi_t}$ and $a^{\phi_{x_i}}$ have one higher order of
$p$ comparing with those in \cite{GSh4} and we have Lemma \ref{Lemm:f}. For the precise relation,
recall that Propositions \ref{Prop:phi_t} and \ref{Prop:phi_x1} in the present paper. 
For this reason, we leave the sketch of the proof in the appendix.

\begin{Proposition}\label{eqn:A}
\begin{equation*}
|A^{\phi_t}|+|A^{\phi_{x_i}}| \leq C(t) P^2(t) \ln^{2/3} P(t).
\end{equation*}
\end{Proposition}
{\sc Proof : } First note that
$$
|A^{\phi_t}|+|A^{\phi_{x_i}}|
\leq C \int^t_0 \int_{\yxtt} \int \frac{f(|a^{\phi_t}|+|a^{\phi_{x_i}}|) \,dp \,dy \,d\tau}{(t-\tau)\sqttyx}. 
$$
Let $\delta \in (0, t]$ and $\epsilon \in (0, 1)$. By Lemma \ref{lemma:a_phi}, we get
\begin{align*}
&\int^{\delta}_0 \int_{1-\epsilon < |\xi| <1} \int \frac{f(|a^{\phi_t}|+|a^{\phi_{x_i}}|) \,dp \,dy \,d\tau}{(t-\tau)\sqttyx}\\
& \quad \leq C(t) P^3(t) \int^t_0 \int^{t-\tau}_{(1-\epsilon)(t-\tau)} \frac{r \,dr \,d\tau}{(t-\tau)\sqrt{(t-\tau)^2-r^2}}
 \leq C(t) P^3(t) \sqrt{\epsilon}.
\end{align*}
Again by Lemma \ref{lemma:a_phi} and (\ref{lemma1}) we get
\begin{align*}
&\int^{t-\delta}_0 \int_{\yxtte} \int \frac{f(|a^{\phi_t}|+|a^{\phi_{x_i}}|) \,dp \,dy \,d\tau}{(t-\tau)\sqttyx}\\
& \quad  \leq C(t) P^{3/2}(t) \int^{t-\delta}_0 \int_{\yxtte} 
  		  			 \frac{e^{1/2}(\tau, y)(1-|\xi|)^{-1/4} \,dy \,d\tau}{(t-\tau)\sqttyx}\\
& \quad  \leq C(t) P^{3/2}(t) \int^{t-\delta}_0 (t-\tau)^{-1} 
  		 	  			   \left(\int_{\yxtte} \frac{(1-|\xi|^2)^{-1/2} \,dy}{(t-\tau)^2-|y-x|^2}\right)^{1/2} \,d\tau
\leq C(t) \ln \frac{t}{\delta} P^{3/2}(t) \epsilon^{-1/4}.
\end{align*}
For the tip of the cone we have by Lemma \ref{lemma:a_phi} that
\begin{align*}
&\int^t_{t-\delta} \int_{\yxtt} \int \frac{f(|a^{\phi_t}|+|a^{\phi_{x_i}}|) \,dp \,dy \,d\tau}{(t-\tau)\sqttyx}\\
&\quad \leq C(t) P^3(t) \int^t_{t-\delta} \int^{t-\tau}_0 \frac{r \,dr \,d\tau}{(t-\tau)\sqrt{(t-\tau)^2-r^2}} 
 = C(t) P^3(t) \delta.
\end{align*}
Collecting all above three estimates, we have
$$
|A^{\phi_t}|+|A^{\phi_{x_i}}| \leq C(t) \big[ P^3(t)\sqrt{\epsilon} + \ln\frac{t}{\delta}P^{3/2}(t)\epsilon^{-1/4}
							  	   +P^3(t) \delta \big].
$$
We take
$\delta = \min\{t,  P^{-1}(t)\}$
then
$$
|A^{\phi_t}|+|A^{\phi_{x_i}}| \leq C(t) \big[ P^3(t)\sqrt{\epsilon}+\ln P(t) P^{3/2}(t)\epsilon^{-1/4} +P^2(t)\big].
$$
Taking
$
\epsilon = P^{-2}(t) \ln^{4/3}P(t)
$
the proposition follows. \hfill $\Box$

\subsection{Proof of Theorem \ref{main_2}}
We note that from Subsection \ref{estimatesDphi} we have proved that $\|D\phi(t)\| \leq C(t)$.
Since $|p|<P(0)$ in the second terms of the representations $\partial_t\phi$
and $\partial_{x_i}\phi$, one can see that these terms are also bounded by $C(t)$.
Now with Propositions \ref{eqn:BC} and \ref{eqn:A} we obtain
$$
|\partial_t \phi(t, x)|+|\nabla_x \phi(t, x)| 
\leq C(t)P^2(t) \ln^{3/2} P(t) + C(t) \int^t_0 \|\partial_t \phi(\tau)\|+\|\nabla_x \phi(\tau)\|\,d\tau.
$$
So for fixed $t$ and $s \in [0, t]$
$$
\|\partial_t \phi(s)\|+\|\nabla_x \phi(s)\| 
\leq C(t)P^2(t) \ln P(t) + C(t) \int^s_0 \|\partial_t \phi(\tau)\|+\|\nabla_x \phi(\tau)\|\,d\tau.
$$
By Gronwall's inequality and taking $s=t$ we achieve
\begin{equation}\label{eqn:phi_final}
\|\partial_t \phi(t)\|+\|\nabla_x \phi(t)\|\leq C(t)P^2(t) \ln P(t).
\end{equation}
We define the characteristics $(\mathcal{X}, \mathcal{P})(s, t, x, p)$ for (\ref{NV2}) by
\begin{align}\label{character1}
d/ds \mathcal{X} &= \hat{\mathcal{P}}\\
d/ds \mathcal{P} &= -(\partial_t\phi(s, \mathcal{X})+\hat{\mathcal{P}}\cdot\nabla_x\phi(s, \mathcal{X}))\mathcal{P} 
	   	  -\frac{\nabla_x \phi(s, \mathcal{X})}{\sqrt{1+|\mathcal{P}|^2}},\label{character2}
\end{align}
with $\mathcal{X}(t, t, x, p)=x$, $\mathcal{P}(t, t, x, p)=p$.
Also define
$$
\bar{P}(t):= \sup \{e^{\phi(s,x)}|p| : f(s, x, p) \neq 0 \text{ for some } (s, x) \in [0, t]\times \mathbb{R}^2\}+3.
$$
Consider $e^{2\phi}|p|^2$ along the characteristics :
$$
d/ds (e^{2\phi(s, \mathcal{X})}|\mathcal{P}|^2) 
			 = -2 e^{2\phi(s, \mathcal{X})}\hat{\mathcal{P}} \cdot \nabla_x \phi(s, \mathcal{X}).
$$
Then with (\ref{eqn:phi_final}) one can see that when $f(0, x, p)\neq 0$,
\begin{align*}
e^{2\phi(0, \mathcal{X}(0, t, x, p))}|\mathcal{P}(0, t, x, p)|^2 
	  	 		&\leq e^{2\phi(t, x)}|p|^2 + C \int^t_0 e^{2\phi(\tau, \mathcal{X})} \|\nabla_x\phi(\tau)\| \,d\tau\\
				& \leq e^{2\phi(t, x)}|p|^2  + C(t) \int^t_0 e^{2\phi(\tau, \mathcal{X})}  P^2(\tau) \ln P(\tau) \,d\tau.
\end{align*}
Note that in (\ref{eqn:phi}) we have $e^\phi \leq e^{\phi_{\rm hom}}$. 
Also we have seen that in Subsection \ref{estimatesDphi}
$\|\phi_{\rm hom}(t)\| \leq C(1+t)$. Therefore with the definition $\bar{P}$, for fixed $t$ and $s \in [0, t]$
we get
$$
\bar{P}^2(s) \leq C(t)+C(t)\int^s_0 \bar{P}^2(\tau)\ln \bar{P}^2(\tau) \,d\tau.
$$ 
Therefore we have
$$
\bar{P}^2(s) \leq \exp(e^{C(t)s} \ln C(t))
$$
and taking $s=t$ we get
\begin{equation}\label{eqn:barP}
\bar{P}(t) \leq C(t).
\end{equation}
Note that by (\ref{eqn:barP}) we have
$$
\int \frac{f(t, x, p)}{\sqp} \,dp 
\leq C(t) \int_{e^\phi |p| \leq C(t)} \big(e^\phi |p|\big)^{-1} d(e^\phi p) \leq C(t).
$$
So in (\ref{eqn:phi}) we get $-\phi \leq C(t)$ and so $e^{-\phi} \leq C(t)$. 
Therefore with the definition $\bar{P}$ and (\ref{eqn:barP}) we achieve
$P(t) \leq C(t).$ \hfill $\Box$

\appendix
\section{Proof of Proposition \ref{Prop:phi_x1}}  
Using the following operations
\begin{align*}
S &:= \partial_t + \hat{p}\cdot \nabla_x\\
T_k &:= \frac{1}{\sqxi}(\partial_{x_k} -\xi_k \partial_t) \quad k=1, 2.
\end{align*}
we obtain
\begin{align}
\partial_t & = \frac{S- \sqxi (\hat{p}_1 T_1 + \hat{p}_2 T_2)}{\xiDp} \notag\\
\partial_{x_1} & = \frac{\xi_1 S + \sqxi [(1+\xi_2 \hat{p}_2)T_1 - \xi_1 \hat{p}_2 T_2]}{\xiDp} \label{def:x1}\\
\partial_{x_2} & = \frac{\xi_2 S + \sqxi [-\xi_2 \hat{p}_1 T_1 + (1+\xi_1 \hat{p}_1)T_2]}{\xiDp}. \notag
\end{align}
By these definitions it follows from (\ref{eqn:phi}) that
\begin{align*}
\partial_{x_1} \phi (t, x) &= \partial_{x_1} \phi_0\\ 
		    & \quad -2 \int^t_0 \int_{\yxtt}\int \frac{\xi_1 Sf(\tau, y, p)}{(\xiDp)\sqp\sqttyx}\,dp \,dy \,d\tau\\
			& \quad -2 \int^t_0 \int_{\yxtt}\int \frac{(1+\xi_2\hat{p}_2) T_1 f(\tau, y, p)}{(\xiDp)\sqp(t-\tau)}\,dp \,dy \,d\tau\\
			& \quad +2 \int^t_0 \int_{\yxtt}\int \frac{\xi_1\hat{p}_2 T_2 f(\tau, y, p)}{(\xiDp)\sqp(t-\tau)}\,dp \,dy \,d\tau\\
			& := \partial_{x_1} \phi_0 + S {\rm term}_{x_1} + T_1 {\rm term}_{x_1} + T_2 {\rm term}_{x_1}.
\end{align*}
Note that from (\ref{NV2})
\begin{equation}\label{eq:Sf}
Sf = F(t, x, p) \cdot \nabla_p f + 3 S (\phi) f
\end{equation}
where $$ F(t, x, p) := S (\phi) p + \frac{\nabla_x \phi}{\sqp}.$$
So $S {\rm term}_{x_1}$ becomes
\begin{align*}
S {\rm term}_{x_1} &= -2 \int^t_0 \int_{\yxtt} \int \frac{\xi_1 F(\tau, y, p) \cdot \nabla_p f(\tau, y, p)}{(\xiDp)\sqp\sqttyx}\,dp \,dy \,d\tau \\
  	   		   & -2 \int^t_0 \int_{\yxtt} \int \frac{3 \xi_1 S(\phi) f(\tau, y, p)}{(\xiDp)\sqp\sqttyx}\,dp \,dy \,d\tau\\
			   & = -2 \int^t_0 \int_{\yxtt} \int \frac{\xi_1 S(\phi) f(\tau, y, p)}{\sqp\sqttyx}\,dp \,dy \,d\tau \\
			   & -2 \int^t_0 \int_{\yxtt} \int \frac{\xi_1 (\xi + \hat{p}) \cdot (\nabla_x\phi)  f(\tau, y, p)}{(\xiDp)^2 (1 +|p|^2)^{3/2}\sqttyx}\,dp \,dy \,d\tau.
\end{align*}
Also note that
\begin{equation}\label{prop:Tk}
\frac{\partial}{\partial y_k}\left(\frac{g(\tau, y)}{\sqxi}\right) 
+ \frac{\partial}{\partial\tau}\left(\frac{-\xi_k g(\tau, y)}{\sqxi}\right) = T_k g(\tau, y).
\end{equation}
So we get
\begin{equation*}
T_1 {\rm term}_{x_1} = - 2 \int^t_0 \int_{\yxtt} \int \frac{(1+\xi_2\hat{p}_2)\nabla_{(\tau, y)}}{(\xiDp)\sqp(t-\tau)}
		 		 \cdot \left(\frac{(-\xi_1, 1, 0)f}{\sqxi}\right) \,dp \,dy \,d\tau.
\end{equation*}
We integrate by parts :
\begin{align} \label{eqn:T1term_xi}
& \int^t_0 \int_{|y-x|<(1-\epsilon)(t-\tau)} \int K_{T_1}\nabla_{(\tau, y)}
		 		 \cdot \left(\frac{(-\xi_1, 1, 0)f}{\sqxi}\right)\,dp \,dy \,d\tau\\
&= - \int^t_0 \int_{|y-x|<(1-\epsilon)(t-\tau)} \int \nabla_{(\tau, y)}K_{T_1} \cdot
   	  		   \left(\frac{(-\xi_1, 1, 0)f}{\sqxi}\right)\,dp \,dy \,d\tau \notag\\
& \quad + \int^t_0 \int_{|y-x|=(1-\epsilon)(t-\tau)} \int K_{T_1} \frac{(-\xi_1, 1, 0)f}{\sqxi} \cdot
  			 \left(1-\epsilon, \frac{y_1-x_1}{|y-x|}, \frac{y_2-x_2}{|y-x|}\right)\,dp \,dS_y \,d\tau \notag\\
& \quad + \int_{|y-x|<(1-\epsilon)t} \int \left. K_{T_1} \frac{(-\xi_1, 1, 0)f}{\sqxi} \right|_{\tau=0} \cdot (-1, 0, 0) \,dp\,dy\notag
\end{align}
where $\epsilon \in (0, 1)$ and 
$$ K_{T_1} := \frac{1+\xi_2\hat{p}_2}{(\xiDp)\sqp(t-\tau)}.$$
In the second term in (\ref{eqn:T1term_xi}) note that
\begin{equation}\label{eqn:epsilon}
\left|\frac{(-\xi_1, 1, 0)}{\sqxi} \cdot \left(1-\epsilon, \frac{y_1-x_1}{|y-x|}, \frac{y_2-x_2}{|y-x|}\right) \right|
= \sqrt{2\epsilon-\epsilon^2}\frac{|y_1-x_1|}{(1-\epsilon)(t-\tau)}\leq \sqrt{2\epsilon}.
\end{equation}
So let $\epsilon \rightarrow 0^+$. We get
\begin{align*}
T_1 {\rm term}_{x_1} &= -2 \int_{\yxt}\int \frac{\xi_1 K_{T_1} f^{\rm in} (y, p)}{\sqxi} \,dp \,dy\\
		 		 & +2 \int^t_0 \int_{\yxtt} \int \nabla_{(\tau, y)}K_{T_1} \cdot
				   \left(\frac{(-\xi_1, 1, 0)f}{\sqxi}\right)\,dp \,dy \,d\tau.
\end{align*}
Note that
\begin{equation*}
\nabla_{(\tau, y)}K_{T_1} \cdot (-\xi_1, 1, 0) = 
-\frac{\xi_1 \xi_2 \hat{p}_2}{(\xiDp)\sqp(t-\tau)^2}-\frac{(\xi_1+\hat{p}_1)(1+ \xi_2\hat{p}_2)}{(\xiDp)^2\sqp(t-\tau)^2}.
\end{equation*}
One can identify $T_2 {\rm term}_{x_1}$ by the same argument. Combining these terms,
we obtain the representation of $\partial_{x_1}\phi$.

\section{Proof of Lemma \ref{Lemm:f}}
In (\ref{eqn:phi}) note that $e^\phi \leq e^{\phi_{\rm hom}}$.
Let $(\mathcal{X},\mathcal{P})(s,t,x,p)$ denote the characteristics as in (\ref{character1}) and (\ref{character2}).
In short, we use $\mathcal{X}(s) := \mathcal{X}(s, t, x, p)$ and $\mathcal{P}(s) := \mathcal{P}(s, t, x, p)$.
Note that the function $e^{-3\phi}f$ is constant along these curves. Hence the solution of (\ref{NV2}) is given by 
\begin{align*}
f(t,x,p)&= f^{\rm in}(\mathcal{X}(0), \mathcal{P}(0))
\exp \left[3\phi(t,x)\right] 
\exp \left[-3\phi^{\rm in}_0(\mathcal{X}(0))\right]\\
&\leq f^{\rm in}(\mathcal{X}(0),\mathcal{P}(0))\exp \left[3\phi_{\rm hom}(t,x)\right]
\exp \left[-3\phi^{\rm in}_0(\mathcal{X}(0))\right].
\end{align*}
So with (\ref{eqn:esti.phi_hom}) the lemma follows.

\section{Proof of Lemma \ref{lemma:a_phi}}
Note that
\begin{align*}
|a^{\phi_t}|+|a^{\phi_{x_i}}| & \leq C (1+|p|)(1-|\hat{p}|^2)(\xiDp)^{-2}[|\xi+\hat{p}|+|\xi\wedge\hat{p}|]\\
							  & \leq C (1+|p|)(1-|\hat{p}|^2)(\xiDp)^{-3/2}.
\end{align*}
Then we have
\begin{equation}\label{help:awithP}
\int (|a^{\phi_t}|+|a^{\phi_{x_i}}|)f \,dp \leq C P(t) \int_{|p|<P(t)}f (1-|\hat{p}|^2)(\xiDp)^{-3/2} \,dp.
\end{equation}
Let $\hat{u}=(1+u^2)^{-1/2}u$. One can see that
\begin{equation}\label{eqn:angular}
\int^{\pi}_0 \frac{\,d\theta}{(1-\hat{u}\cos\theta)^{3/2}} 
\leq \frac{1}{\sqrt{1-\hat{u}|\xi|}} \int^{\pi}_0 \frac{\,d\theta}{1-\hat{u}|\xi|\cos\theta}
= \frac{1}{\sqrt{1-\hat{u}|\xi|}} \frac{\pi}{\sqrt{1-\hat{u}^2|\xi|^2}} 
\leq C (1-\hat{u}|\xi|)^{-1}.
\end{equation}
So for any $R \in (0, P(t)]$ using (\ref{eqn:angular}) we get
\begin{equation}\label{eqn:forR}
\int_{|v|<R} (1-|\hat{p}|^2)(\xiDp)^{-3/2}\,dp 
\leq C \int^R_0 (1+\hat{u}) u \,du
\leq C \int^R_0 u\,du \leq CR^2.
\end{equation}
If $R=P(t)$, then with (\ref{help:awithP}) we get 
\begin{equation}\label{eqn:RP(t)}
\int (|a^{\phi_t}|+|a^{\phi_{x_i}}|)f \,dp \leq C(t) P^3(t).
\end{equation}
For $R < P(t)$ we use H\"older's inequality and (\ref{eqn:angular}) to obtain
\begin{align*}
&\int_{R<|p|<P(t)} f (1-|\hat{p}|^2)(\xiDp)^{-3/2}\,dp \\
& \quad \leq \int_{R<|p|<P(t)} 2 f (\xiDp)^{-1/2} \,dp 
\leq 2 \left(\int_{R<|p|} f^{3/2} \,dp \right)^{2/3} \left( \int_{|p|<P(t)}(\xiDp)^{-3/2} \,dp\right)^{1/3}\\
& \quad \leq C \left((1+R^2)^{-1/2}\int_{R<|p|} f \sqp \,dp \right)^{2/3}
  			 \left(\int^{P(t)}_0 (1-u|\xi|)^{-1} u \,du \right)^{1/3}\\
& \quad \leq CR^{-2/3}e^{2/3}\left(\int^{P(t)}_0 (1-|\xi|^2)^{-1}u\,du\right)^{1/3}
\leq C R^{-2/3}e^{2/3}P^{2/3}(t) (1-|\xi|^2)^{-1/3}.
\end{align*}
Now combining this and (\ref{eqn:forR}) we get
$$
\int (|a^{\phi_t}|+|a^{\phi_{x_i}}|)f \,dp \leq C(t)P(t) (R^2+R^{-2/3}e^{2/3}P^{2/3}(t)(1-|\xi|^2)^{-1/3})
$$
for all $R \in (0, P(t))$.
Taking
$R = e^{1/4}P^{1/4}(t)(1-|\xi|^2)^{-1/8}$
the lemma follows with (\ref{eqn:RP(t)}).


\begin{thebibliography}{99} 
 
\bibitem{C}  
S.~Calogero: {\it Spherically symmetric steady states of galactic dynamics in scalar gravity.}  
Class. Quant. Gravity {\bf 20}, 1729--1741 (2003)
 
 
\bibitem{C2} S.~Calogero: {\it Global Small Solutions of the Vlasov-Maxwell System in the Absence of Incoming Radiation.}
Indiana Univ. Math. Journal (to appear), Preprint: math-ph/0211013 
 
\bibitem{CaLe} S.~Calogero, H.~Lee: {\it The non-relativistic limit of the Nordstr\"om-Vlasov system.}
Preprint: math-ph/0309030

\bibitem{CaRe}  
S.~Calogero, G.~Rein: {\it On classical solutions of the Nordstr\"om-Vlasov system.} Comm. Partial Diff. Eqns. 
(to appear), Preprint: math-ph/0304021 

\bibitem{CaRe1}
S.~Calogero, G.~Rein: {\it Global weak solutions to the Nordstr\"om-Vlasov system.} Preprint: math-ph/0309046.
 
\bibitem{DL} R.~DiPerna, R.~J.~Lions: {\it Global weak solutions of Vlasov-Maxwell systems.} Comm. Pure Appl. Math. {\bf 42}, no. 6,
729--757 (1989)

\bibitem{EF} A.~Einstein, A.~D.~Fokker: {\it Die Nordstr\"omsche Gravitationstheorie vom Standpunkt des absoluten Differentialkalk\"uls.}
Annalen der Physik {\bf 44}, 321--328 (1914)

\bibitem{GSh1} R.~Glassey, J.~Schaeffer: {\it Global existence for the relativistic 
Vlasov-Maxwell system with nearly neutral initial data.} Comm. Math. Phys. {\bf 119}, 353--384 (1988) 
 
\bibitem{GSh2} R.~Glassey, J.~Schaeffer: {\it The ``Two and One-Half Dimensional" Relativistic Vlasov-Maxwell System.} 
Comm. Math. Phys. {\bf 185}, 257--284 (1997)

\bibitem{GSh3} R.~Glassey, J.~Schaeffer: {\it The relativistic Vlasov-Maxwell system in two space dimensions, Part I.}
Arch. Rational Mech. Anal. {\bf 141}, 331--354 (1998)

\bibitem{GSh4} R.~Glassey, J.~Schaeffer: {\it The relativistic Vlasov-Maxwell system in two space dimensions, Part II.}
Arch. Rational Mech. Anal. {\bf 141}, 335--374 (1998)


\bibitem{GS1} R.~Glassey, W.~Strauss: {\it Singularity formation in a 
collisionless plasma could occur only at high velocities.} Arch. Rational Mech. Anal. {\bf 92}, 59--90 (1986) 
 
\bibitem{GS2} R.~Glassey, W.~Strauss: {\it Absence of shocks in an initially dilute 
collisionless plasma.} Comm. Math. Phys. {\bf 113}, 191--208 (1987)  
 
\bibitem{LE} H.~Lee: {\it The classical limit of the relativistic Vlasov-Maxwell system in two space dimensions.}
Math. Methods Appl. Sci. (to appear)

\bibitem{LP} 
P.-L.~Lions, B.~Perthame: {\it Propagation of moments and regularity for the 3-dimensional 
Vlasov-Poisson system.} Invent. Math. {\bf 105}, 415--430 (1991) 

\bibitem{No}  
G.~Nordstr\"om:  {\it Zur Theorie der Gravitation vom Standpunkt des Relativit\"atsprinzips.}  
Ann. Phys. Lpz. {\bf 42}, 533 (1913)   

\bibitem{Pf} 
K.~Pfaffelmoser: {\it Global classical solutions of the Vlasov-Poisson system in three 
dimensions for general initial data.}  J. Diff. Eqns. {\bf 95}, 281--303 (1992) 

\bibitem{R} 
G.~Rein: {\it Selfgravitating systems in Newtonian theory---the Vlasov-Poisson system.} 
Banach Center Publications {\bf 41}, Part I, 179--194 (1997) 

\bibitem{R2} G.~Rein: {\it Generic global solutions of the relativistic 
Vlasov-Maxwell system of plasma physics.} Comm. Math. Phys. {\bf 135}, 41--78 (1990) 
 
\bibitem{RR1} 
G.~Rein, A.~D.~Rendall: {\it Global existence of solutions of the spherically symmetric 
Vlasov-Einstein system with small initial data.} 
Commun. Math. Phys. {\bf 150}, 561--583 (1992) 

\bibitem{Sch2} 
J.~Schaeffer: 
{\it Global existence of smooth solutions to the Vlasov-Poisson system 
in three dimensions.} Comm. Partial Diff. Eqns. {\bf 16}, 1313--1335 (1991) 
  
\end{thebibliography}
\end{document}